\documentclass[11pt,twoside]{article}

\usepackage{asp2006}
\usepackage{epsf}
\usepackage{epsfig}
\usepackage{lscape}
\usepackage{graphicx}
\begin{document}
\title{Systematic study of a large sample of NLS1 galaxies from SDSS:
first results}

\author{W. Yuan\altaffilmark{1}, H. Zhou\altaffilmark{2},
T. Wang\altaffilmark{2},
X. Dong\altaffilmark{2},
H. Lu\altaffilmark{2},
J. Wang\altaffilmark{2},
Y. Lu\altaffilmark{2},
L. Fan\altaffilmark{2},
L. Dou\altaffilmark{1}
}
\affil{}    

\altaffiltext{1}{Yunnan Observatory, National Astronomical
Observatories, Chinese Academy of Sciences, P.O.
Box 110, Kunming, Yunnan 650011, China; wmy@ynao.ac.cn}
\altaffiltext{2}{Center for Astrophysics, University of Science and
Technology of China, Hefei, Anhui, 230026, China;
mtzhou,twang@ustc.edu.cn}

\begin{abstract}
This proceedings contribution summarizes first results
obtained from a systematic study of 2011 NLS1 galaxies,
 as presented in  Zhou et al.\ (2006, ApJS 166, 128).
The sample was compiled
by examining the spectral parameters of galaxies and QSOs
derived from SDSS DR3 data.
We discuss some preliminary results on the statistic properties
of the sample,
such as the fraction of NLS1,
the properties of broad and narrow emission lines,
and emission in other wavebands. The black hole mass -- velocity
dispersion relation for NLS1s was re-examined using the velocity
dispersion values estimated from the stellar absorption spectra of
the host galaxies.
Preliminary result from an X-ray study for a small subset
using data obtained by XMM-Newton is briefly discussed.
\end{abstract}


\section{Introduction}
Narrow line Seyfert 1 galaxies (NLS1s) are a special type of
active galactic nuclei (AGN)
whose optical spectra are similar to Seyfert\,1 type,
but with much narrower Balmer line width;
they also show weak [OIII] lines and strong FeII emission
in optical and UV (Osterbrock \& Pogge 1985).
The importance of NLS1s lies in that
their properties are at the extremity among
Seyfert\,1 family, including steep soft X-ray
spectra ($E<2\,keV$) and rapid X-ray variability,
strong blue wings of emission lines
(e.g.\ Boroson \& Green 1992; Laor et al. 1994; Wang 1996,
Boller et al. 1996, Grupe et al.\ 1999, 2001,
Xu, et al.\ 2003).
Observational data show that NLS1 tend to have small
black hole masses and the high Eddington ratios $L/L{\rm Edd}$
(e.g.\ Boroson 2002).
In fact, they were found to locate at one extreme end
of eigenvector\,1 of the correlation matrix
(Boroson \& Green 1992, Sulentic et al.\ 2000),
which is believed to be driven primarily by $L/L_{\rm Edd}$.
Radio-loud NLS1s were also found to exist,
though their number is still small
(e.g. Komossa et al.\ 2006).

Yet there are quite a few aspects about NLS1s
where controversies
remain or little study has been given to. The reasons
can be ascribed partly  to the relatively small number of
NLS1 known (a few hundreds) before this work, as well as
the  heterogeneity of the objects as a whole.
For example, little work has been done regarding
broad band SED,
associated outflows,  radio-loud NLS1s and their jets,
host galaxies, optical variability, mean spectrum
and dispersion,
the $M_{\rm BH}-\sigma$ relation,
as well as their cosmic evolution.
A much larger and homogeneously selected NLS1 sample is
required. With its large sky coverage and unprecedented
sensitivity and uniformity, the spectroscopic survey of the
Sloan Digital Sky Survey (SDSS; York et al.\ 2000)
provides
us with the most suitable database for discovering NLS1s.
As a pioneer work, Williams et al. (2002) have compiled a
sample of ~150 NLS1 from the SDSS Early Data Release.

We carried out an extensive and systematic search for
NLS1s by making use of the SDSS DR3 data, which
yielded 2011 NLS1s, mostly previously unknown.
The sample and preliminary statistical results
are presented in Zhou et al.\ (2006).
In this proceedings contribution, we summarize
the sample properties and some of the significant
results, with some  new, preliminary development.

\section{Sample and statistical results}
\subsection{Sample}
Our NLS1s were drawn from the SDSS DR3 spectroscopic databases
of both galaxies and QSOs. The data reduction
procedure, as described in detail in Zhou et al.\ (2006), consists
of mainly two steps. Firstly, star light contribution to fiber
spectra from host galaxies is carefully removed by performing
starlight-nuclear spectral decomposition using our algorithm (Lu et
al. 2006) developed at University of Science of Technology of China.
Our technique has proved to be accurate and robust when
performing extensive testing against results from independent
observations and measurement, including HST spectroscopic
measurements (Wang T., et al.\ in prep.). Secondly,
 nuclear emission-line
spectra---with the star-spectrum and  continuum subtracted---are
fitted to derive spectral parameters using the code as
described in Dong et al.\ (2005). In this process, the Balmer
emission lines are de-blended into a broad and a narrow
component---this is essential as by doing so we could really measure
the physical parameters of the broad line region. This process
 accounts for the difference in the results for some of the
objects between our work and that of  Williams et al.\ (2002).

We selected NLS1s as those having FWHM of the {\em broad} component
of H$\beta$ or H$\alpha$ (detected at the $>10~\sigma$ confidence
level) narrow than  $2200~km~s^{-1}$. The 10\% relaxed margin in the
linewidth cutoff compared to the conventional $2000~km~s^{-1}$
(Goodrich 1989) is to
recover those NLS1s dropped out due to the uncertainty in linewidth
measurement.
Our criteria resulted in 2011 NLS1 candidates.  Using a stricter
criterion of $2000\,km\,s^{-1}$ would reduce the sample to
1885 objects.
All but one fulfill $\rm [OIII]/H\beta<3$---the second item of the
conventional definition. We gave, along with the source
list\footnote{A table of the sample including measured spectral
parameters is available as on-line data associated with the paper
 Zhou et al.\ (2006, ApJS 166, 128).},
measured spectral parameters and their error estimates.
 Below we summarize some of the preliminary
results directly based on statistical analysis of observed and
derived parameters; for details of these analysis please refer to
Zhou et al.\ (2006).

\subsection{Summary of statistical results}

\noindent{1) NLS1 fraction}\\
We found a strong dependence of the NLS1 fraction among all
Seyfert\,1 AGNs.  With increasing optical luminosity, the fraction
of NLS1s first increases and peaks at $M_{g'}\sim-22$, and then
drops again rapidly.
The NLS1 fraction also depends on radio-loudness,
dropping from $\sim$15\% in radio-quiet
to only 5\% in radio-loud objects. The NLS1 fraction in optically
selected AGN sample is much lower than that in soft X-ray (ROSAT)
selected AGN samples.

\noindent{2) Emission line properties}\\
On average the relative  FeII emission,
$R_{4570}=FeII(4434-4684)/H\beta$, in NLS1s is about twice
that in normal AGNs, and is anti-correlated with the broad component
width of the Balmer emission lines.
The equivalent width of H$\beta$ and FeII
emission lines are strongly positively  correlated with
the H$\beta$ and continuum luminosities
(the 'inverse' Baldwin effect).
We do not find any difference
between NLS1s and normal AGNs
in regard to the Narrow Line Region.

\noindent{3) $M_{\rm BH}$--$\sigma$} relation\\
We  have examined the black hole  mass vs.\
stellar velocity dispersion ($\sigma_{*}$) relation for
a subsample of  308 NLS1s
for which $\sigma_{*}$ could be
measured directly from fitting the starlight in the SDSS spectra
with our stellar spectral templates. A significant correlation
between $M_{BH}$ and $\sigma_{*}$ is found,
but with the bulk of  black hole masses
falling below the values
expected from the $M_{BH}-\sigma_{*}$ relation for normal galaxies
and normal AGNs (Figure\,1, left panel).
This result indicates that NLS1s are underage AGNs,
where the growth of the SMBH lags behind the formation of the
galactic bulge
(see also, e.g.\ Mathur et al.\ 2001, Grupe et al. 2004,
Bian \& Zhao 2004, Botte et al.\ 2004).
We also find that the width of [NII] is well
correlated with $\sigma_{*}$.

\noindent{4) X-ray properties}\\
635 NLS1s in the sample have X-ray
counterparts detected in the RASS.
The well-known anti-correlation between the width of
broad low-ionization lines and the soft X-ray spectral slope
for broad line AGNs extends down to $FWHM\sim
1~000~km~s^{-1}$ in NLS1s, but the trend appears to reverse at still
smaller line widths.
There are 28 NLS1s being observed by XMM-Newton
either as targets or serendipitously, for which their
spectra were analyzed.
For most of the objects the 0.3--10\,keV X-ray continuum
can be fitted with a power law plus a soft X-ray excess,
mostly well parameterized by a blackbody component.
No significant Fe K$\alpha$ line was  detected.
In  Figure\,1 (right panel)
we plot the distribution of the photon indices
of the underlying continuum derived from XMM-Newton data.
The spectral slopes are systematically steeper than those for
broad line Seyfert 1 galaxies, confirming previous results
(e.g.\ Leighly 2000).

\begin{figure}[tbp]
\plotone{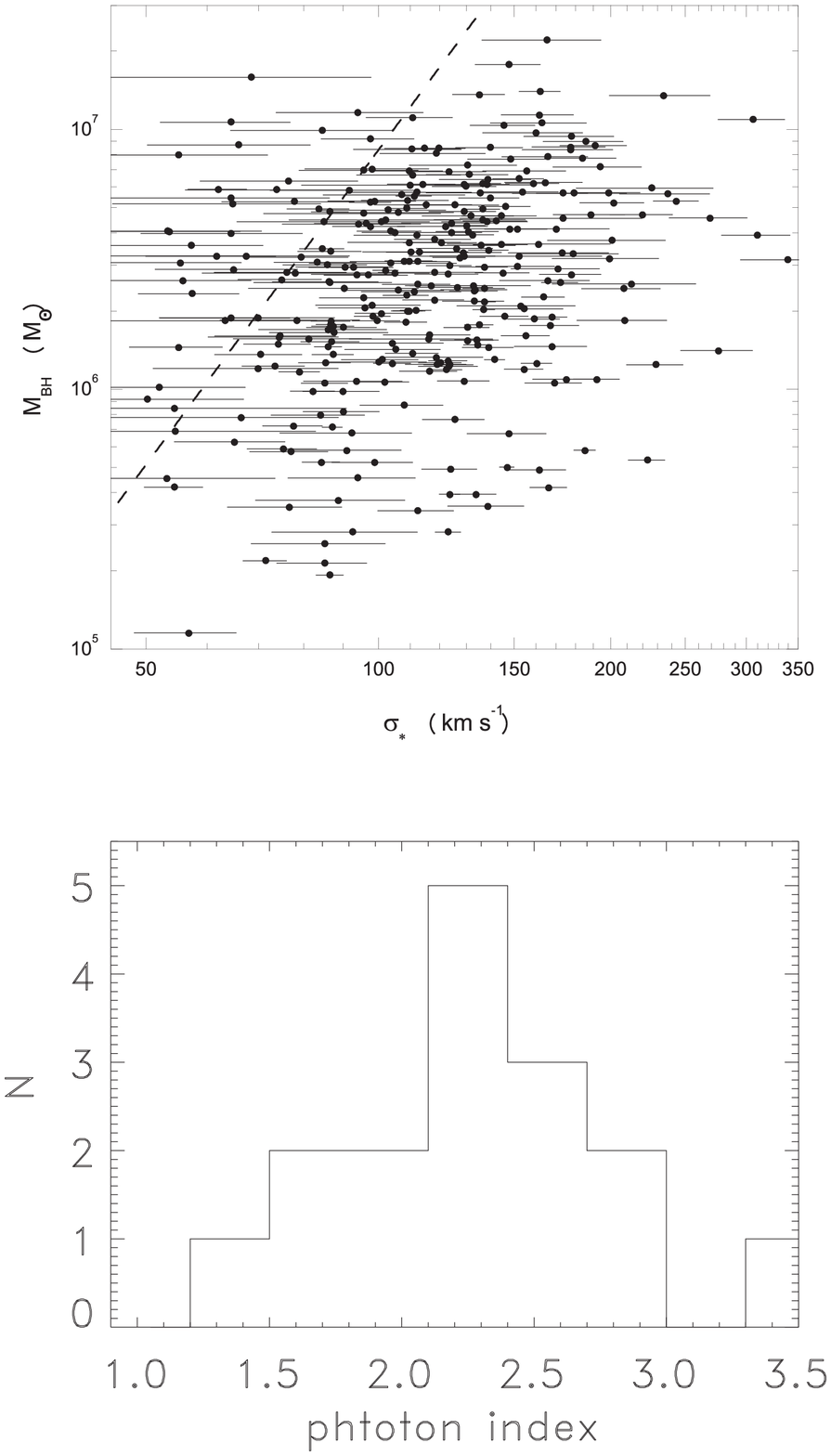} \caption{Upper panel: The $M_{\rm
BH}$--$\sigma$ relation for 308 NLS1s for which $\sigma_{*}$ is
reliably measured, with the dash line as the Tremaine et al.\ (2002)
relation for normal galaxies. Lower panel: Distribution of the
photon indices of the underlying 0.3--10\,keV X-ray continuum for
those having XMM-Newton data}
\end{figure}

\noindent{5) Radio-loud NLS1s}\\
Of the sample 142 (7\%) NLS1s were detected  in radio
in the FIRST survey.
About two dozen
are very radio loud (radio-loudness $>$100, Zhou et al.\ in prep.).
Of particular interest, some objects show hybrid properties of
NLS1 and blazar---with 2MASX\,J0324+3410 being an extreme example
(Zhou et al.\ 2007, ApJ in press).
A detailed investigation of radio-loud fraction of NLS1s is underway,
by taking into account
various selection effects in the SDSS spectroscopic survey.

\subsection{On-going and future work}
We are currently working on updating the NLS1 AGN sample
using the latest SDSS DR5 data.
Other work currently undertaken includes:
analyzing the X-ray spectral and temporal data for
those having XMM-Newton and ROSAT observations,
examining optical variability of NLS1 using
the SDSS supernovae survey data (Ai et al.\, this proceedings),
radio-loud NLS1s, broad-band SED.
As for the future, we also plan to work on, as example,
NLS1s' host galaxies and their stellar populations and
comparisons with those of broad line Seyfert AGNs,
high-redshift NLS1s.



\end{document}